\documentclass[11pt]{article}
\usepackage[letterpaper]{geometry}
\usepackage{url}
\usepackage{latexsym,amssymb,amsmath}
\usepackage{amsfonts,amscd,color,epsfig}
\usepackage{complexity}
\begin{document}

\title{Can quantum mechanics help distributed computing?}
\vspace{15pt}
\author{Anne Broadbent  and   Alain Tapp\\[.2cm]
\normalsize\sl D\'epartement d'informatique et de recherche op\'erationnelle\\[-0.1cm]
\normalsize\sl Universit\'e de Montr\'eal, C.P.~6128, Succ.\ Centre-Ville\\[-0.1cm]
\normalsize\sl Montr\'eal (QC), H3C 3J7~~\textsc{Canada}\\[.2cm]
{IQC, University of Waterloo}
\url{{broadbea, tappa}@iro.umontreal.ca} }

\author{Anne Broadbent \and Alain Tapp}

\date{}

\maketitle

\begin{abstract}

We present a brief survey of results where quantum information
processing  is useful to solve distributed computation tasks. We
describe problems that are impossible to solve using classical
resources but that become feasible with the help of quantum
mechanics. We also give examples where the use of quantum
information significantly reduces the need for communication. The
main focus of the survey is on communication complexity but we also
address other distributed tasks.

\smallskip

\end{abstract}

\section*{Quantum computation and entanglement}

This survey is aimed at researchers  having very limited knowledge of quantum
computation, but that have a basic understanding of complexity from the theoretical computer science perspective. We address the topics of communication complexity and
pseudo-telepathy, as well as other problems of interest
in the field of distributed computation. The goal of this survey is
not to be exhaustive but rather to cover many different aspects and
give the highlights and intuition into the power of distributed
quantum computation. Other relevant surveys are
available~\cite{SteanevanDam,Brassard03, BR03, BBT05}.

In classical computation, the basic unit of information is the bit.
In quantum computation, which is based on quantum mechanics, the
basic unit of information is the {\em qu}bit. A string of bits can
be described by a string of zeroes and ones; quantum information can
also be in a classical state represented by a binary string, but in
general it can be in {\em superposition} of all possible strings
with different {\em amplitudes}. Amplitudes are complex numbers and
thus the complete description of a string of $n$~qubits
requires~$2^n$ complex numbers. The fact that quantum information
uses a continuous notation does not mean that qubits are somewhat
equivalent to analog information: although the description of a
quantum state is continuous, quantum measurement, the method of
extracting classical information from a quantum state, is discrete.
Only $n$ bits of information can be extracted from an $n$-qubit
state. Depending of the choice of measurement, different properties
of the state can be extracted but the rest is lost forever. Another
way to see this is that measurement disturbs a quantum state
irreversibly. In quantum algorithms, it is possible to compute a
function on all inputs at the same time by only one use of a quantum
circuit. The difficult part is to perform the appropriate
measurement to extract useful information about the function.
We refer the reader to \cite{Chuang, Mermin, KLM07}
for introductory textbooks to quantum information processing.

One of the most mysterious manifestations of  quantum information is
\emph{entanglement}, according to which distant parties can share
correlations that go beyond what is feasible with classical
information alone: \emph{quantum correlations}! Entanglement is
strange, useful and not completely understood. Some of the results
described in this survey will shed light on this  facet of quantum
mechanics. In the absence of quantum entanglement, it is necessary to transmit $n$~qubits to
convey $n$~bits of information~\cite{Holevo}. When the players share
entanglement, this can be improved to~$2n$ but not more~\cite{IP}. One would therefore think that quantum mechanics cannot
reduce the amount of  communication required in distributed tasks
(by more than a constant). Surprisingly, this intuition is wrong!

We are beginning to get the idea that classical information and
quantum information are quite different. As further evidence, note
that classical information can trivially be copied, but quantum
information is disturbed by observation and therefore cannot be
faithfully copied in general.  However, the fact that quantum
information cannot be copied does not imply that it cannot be
teleported~\cite{teleport}.

Since quantum information cannot, even in theory, be copied, and
since it is very fragile in its physical implementations, it was
initially  believed  by some
that errors would be an unsurmountable barrier to building a quantum
computer. Actually, this was the first and only serious theoretical
threat to quantum computers. Fortunately, quantum error correction
and fault tolerant computation were shown to be possible with
realistic assumptions if the rate of errors is not too big. This
implies that a noisy quantum computer can perform an arbitrary long
quantum computation efficiently as soon as some threshold of gate
quality is attained~\cite{AGJ06}. 

Quantum key distribution (QKD) \cite{BB84} is one of the founding
results of quantum information processing.  This amazing
breakthrough is an amplification protocol for private shared keys.
Another result that propelled quantum computation into the
attractive area of research that it is today is Peter Shor's
factoring algorithm \cite{Shor}, which is  a polynomial-time
algorithm to factor integers on a quantum computer. Note that the
best known classical algorithm, the number field sieve,
\cite{NFSieve, LenstraLenstra} takes time in $O(2^{c n^{1/3}(\log
n)^{2/3}})$ where $n$ is the number of bits of the number to be
factored.  The importance of this result is evidenced by the fact
that the security of most sensitive transactions on the Internet is
based on the assumption that factoring is difficult~\cite{RSA}. We will not discuss quantum
computer implementations but let us mention that  experiments are
only in their infancy. Quantum communication is the most successful
present-day implementation, with QKD being implemented by dozens of
research groups and
being commercially available~\cite{idQuantique}.  

We now begin a survey of the main results in distributed
computation, starting with  detailed reviews of pseudo-telepathy and of communication complexity, followed by an overview of other distributed quantum tasks.  We will not give the quantum algorithms or protocols
that solve the presented problems; they are usually quite simple.
Most of the time, the difficulty is to provide a proof of their
correctness or to show that a classical computer cannot be as
efficient.

\section*{Pseudo-telepathy}

\looseness-1 The term \emph{pseudo-telepathy}   originates from the authors
of~\cite{BCT99} (although it does not appear in the paper). It
involves the study of a phenomenon that
previously appeared in physics literature~\cite{GHZ, MerminGHZ}. We introduce this
strange behaviour of quantum mechanics with~a~story.

Alice and Bob claim that they have mysterious powers that enable
them to perform telepathy. However surprising that this may seem,
they are willing to prove their claim to a pair of physicists that
do not know about quantum mechanics. Imagine that they are willing
to bet a substantial amount of money. To be more precise, Alice and
Bob do not claim that they can send emails by thought alone, but
they claim that they can win the following game with certainty
without talking to each other. As you will see,  their claim is very
surprising because it appears that it is impossible to satisfy!

A magic square (see Figure~\ref{fig:magic-square}) is a 3~by~3
matrix of binary digits such that the sum of each row is even and
the sum of each column is odd.  A simple parity argument is
sufficient to convince oneself that a magic square cannot exist:
since the sum of each row is even, the sum of the whole square has
to be even. But since the sum of each column is odd, the sum of the
whole square has to be odd. This is a contradiction and therefore
such a square cannot exist.

\begin{figure}[ht]
\begin{center}
\begin{tabular}{|c|c|c|}
\hline
 0 & 1 & 1  \\
\hline
 1 & 1 & 0  \\
\hline
 0 & 1 & ?  \\
\hline
\end{tabular}
 \caption{\label{fig:magic-square}A partial magic square. In a magic
square, the sum of each row is even and the sum of each column is
odd.}
\end{center}
\end{figure}

In the game that Alice and Bob agree to play, they will behave
exactly as if they actually agreed on a collection of such squares
(at least, in a context where they cannot talk to each other). The
physicists will prevent Alice and Bob from communicating during the
game; an \emph{easy} solution is to place Alice and Bob several
light years away. According to relativity, any message they would
exchange would take several years to arrive.

To test the purported telepathic abilities, each physicist is paired
with a participant. They then ask simultaneously questions: Alice is
asked a give a row of the square (either row 1, 2~or~3) and  Bob is
asked to give a column (either column 1, 2~or~3). Each time the
experiment is performed, Alice and Bob claim to used a different
magic square. After a certain number of repetitions, the physicists
get together and verify that the sum of each row is even and the sum
of each column is odd, and  moreover that the bit at the
intersection of the row and column is the same. It is not so
difficult to see that if Alice and Bob do not communicate after the
onset of the game,  there is no strategy that wins this game with
probability more than $8/9$. This is the outcome that the pair of
physicists would expect. Instead, they are astounded to see that
Alice and Bob always win, no matter how many times they repeat the
game! 
Alice and Bob have managed to win their bet and accomplish a task
that provably  requires communication, but without communicating!
Hence the name {\em pseudo-telepathy}. How is this possible? Thanks
to quantum mechanics, Alice and Bob can win with probability~1. In
addition to agreeing on a strategy before the experiment, Alice and
Bob share enough entangled particles. If you think winning such a
game is amazing, then now you understand a bit more why we consider
entanglement to be such a wonderful and strange resource. This
simple thought experiment has very important consequences on our
understanding of the world in which we live, both in the physical
and philosophical perspectives~\cite{EPR35, Bell, CHSH}.

More formally, a pseudo-telepathy game is a distributed $k$-player
game where the players can agree on a strategy and can share
entanglement. While the players are not allowed to communicate,
each player is asked a question and should provide an answer. The
game must be such that quantum players can win with probability~1
but classical players cannot. The example we presented comes from~\cite{Aravind}. We refer the reader to a survey specifically on this
subject~\cite{BBT05}.

\section*{Communication complexity}

Communication complexity is the study of the amount of communication
required to compute functions on distributed data in a context of
honest cooperating players. It was first introduced by Harold
Abelson \cite{Abelson} and given in its current form by Andrew Yao
\cite{YaoCC}. A good reference on \emph{classical} communication
complexity is \cite{KushilevitzNisan}. There are several variations
of the basic model; here, we concentrate on the most natural one.
Let~$F$ be a $k$-input binary function. We are in a context where
the $k$~players each have one of the inputs to the function. The
\emph{probabilistic} communication complexity is the amount of bits
that have to be broadcast by the players in order for player number
one to be able to compute~$F$ with probability at least~$2/3$ (in
the worst case). We assume that the players share some random bits
and that they cooperate. The value~$2/3$ is arbitrary and can be
very efficiently improved by parallel repetition. Note that in this
model, we do not care about the computational complexity for every
player, but in general the computation required by the players is
polynomial. The trivial solution that works for all functions is for
each player (except the first one) to broadcast his input. We will
see that sometimes, but not always,  the players can do much better.

Let us illustrate the concept with a simple example. Suppose we have
two players, Alice and Bob, who each have a huge electronic file and
they want to test if these are identical. More formally, they want
to compute the equality function. If one insists that the
probability of success be~1, then Bob has to transmit his entire
file to Alice: any solution would require an amount of communication
equal to Bob's file size. Obviously, if we are willing to tolerate
some errors, there is a more efficient solution. Let~$x$ be Alice's
input and $y$ be Bob's, and assume Alice and Bob share~$z$, a random
string of the same length as $x$ and~$y$. If $x=y$, obviously $x
\cdot z= y \cdot z$ but it is not too hard to see that if $x \neq
y$,  the probability that $x \cdot z= y \cdot z$ is exactly~$1/2$
(here, $x \cdot z$ is taken to be the \emph{binary} inner product:
the inner product of~$x$ and~$z$, modulo~2). In order for Alice to
learn this probabilistic information,  Bob only has to send one bit.
By executing this twice, we have that the function can be computed
correctly with probability~$3/4$.

One might argue that we are cheating by allowing Alice and Bob to
share random bits and not counting this in the communication cost.
We have decided to concentrate on this model since it is natural to
compare it to the quantum case where players also have access to shared entanglement. Also, in general, if Alice and Bob
do not share randomness, they can obtain the same result only with
an additional~$\log n$ bits of communication \cite{Newman91}.

Yao is also responsible for pioneering work in the area of
\emph{quantum} communication complexity~\cite{Yao93}, in which he
asked the question: what if the players are allowed to communicate
qubits (quantum information) instead of classical bits. No answer to
this question was initially advanced. In \cite{CB97}, Richard Cleve
and Harry Buhrman introduced a variation on the model, for which
they showed a separation between the classical and quantum models:
the players communicate classically but they share entanglement
instead of classical random strings. This time, the goal is to
compute the function with certainty. They exhibited a function (more
specifically, a \emph{relation}, also called a \emph{promise
problem}) for three players such that in the broadcast model, any
protocol that computes the function requires~3 bits of
communication. In contrast, if the players share entanglement, it
can be computed exactly  with only~2 bits of classical
communication. The function they studied is not very interesting by
itself but the result is revolutionary: we knew that entanglement
cannot replace communication, and what this result shows is that
entanglement can be used to reduce communication in a context of
communication complexity.

Harry Buhrman, Wim van~Dam, Peter H\o yer and Alain Tapp~\cite{BDHT}
improved the above result by exhibiting a $k$-player function (again
with a promise) such that the communication required for computation
with probability~1 is in $\Theta(k \log{k})$, but if the players
share quantum entanglement, it is in  $\Theta(k)$. They also showed
that it is possible to substitute the quantum entanglement for
quantum communication, resulting in a protocol still with $O(k)$
communication. This was the first non-constant gap between quantum
and classical communication complexity. Once more, the function that
was studied is not natural.

Quantum teleportation~\cite{teleport}, shows that two classical bits of communication, coupled with entanglement, enable the transfer of a qubit. Applying this, we get that any two-player protocol using
quantum communication can be simulated by a protocol using
entanglement and classical communication, at the cost of only
doubling the communication.

The first problem of practical interest where quantum information was shown to be very useful is the appointment scheduling problem. For this problem,
Alice has an appointment calendar that, for each day, indicates
whether or not she is free for lunch. Bob also has his own calendar,
indicating whether or not he is free.  The players
wish to know if there is a day where they are both free for a lunch
meeting. In the classical model, the amount of communication
required to solve the problem is in~$\Theta(n)$. In the quantum
model, this was reduced to $O(\sqrt{n}\log{n})$ in~\cite{BCW99}, and
further improved to $O(\sqrt{n})$ in~\cite{AA05}.

The first exponential separation between classical and quantum
communication complexity was presented in \cite{BCW99} but it was in
the case where the function must be computed exactly.
Later, Ran Raz gave an
exponential separation in the more natural probabilistic model that we have presented,
 but for a contrived problem~\cite{Raz99}. See also
related work~\cite{Gavinsky08}. Note that not all functions can be
computed more efficiently using quantum communication or
entanglement; this is the case of the binary inner
product~\cite{IP}.

\section*{More distributed quantum tasks}

\subsection*{Fingerprinting}

This interesting result was introduced in the context of
communication complexity but is of general interest. It was shown in
\cite{Fingerprinting} that to any bitstring or message, a unique and
very short (logarithmic) quantum fingerprint can be associated.
Although the fingerprint is very small and generated
deterministically, when two such fingerprints are compared, it is
possible to determine with high probability if they are equal. The
concepts of quantum fingerprinting were used in the context of
quantum digital signatures~\cite{GC01}.

\subsection*{Coin tossing}

Moving to a more cryptographic context, one of the simplest and most
useful primitives is the ability to flip coins fairly in an
adversarial scenario. \emph{Strong coin tossing} encompasses the
intuitive features of such a protocol: it  allows $k$~players to
generate a random bit with no bias  (or an exponentially small one),
where \emph{bias} is the notion of a player being able to choose the
outcome. The trivial method of allowing a single player to flip a
coin and announce the result is biased: the player could choose the
outcome to his advantage.

It is possible to base the fairness of a coin toss on computational
assumptions: this is due to the fact that bit commitment can be used
to implement coin toss and that bit commitment itself can be
implemented with computational assumptions~\cite{Crepeau}.
However, we know that when quantum computers become available, some of the assumptions on which these protocols are based will unfortunately
become insecure.
Is there a way to
implement a coin toss using quantum information? It was shown by
Andris Ambainis \cite{Ambainis} that if two players can use quantum
communication, this task can be approximated to some extent without
computational assumptions. If both Alice and Bob are honest, the
coin flip will be fair, otherwise one player can bias the coin toss
by 25\% but no more. This is almost tight since it was proven that
in this context, the bias cannot be reduced lower than approximately
21\% (this result is due to unpublished work of Alexei Kitaev;
see~\cite{GW07} for a conceptually simple proof). Recent work by Andr\'e Chailloux and Iordanis Kerenidis~\cite{CK09} establishes an optimal coin tossing protocol.  Kitaev's lower bound
discouraged quantum cryptographers but it was misleading. In a
context where the coin toss is used to choose a winner (a very
natural application), then we know in which direction each player is
trying to bias the coin toss. Surprisingly, in this context,  called \emph{weak} coin tossing, quantum
protocols exist that have arbitrarily small bias~\cite{Mochon}. See
also~\cite{BBBG08} for a loss-tolerant protocol.

\subsection*{Quantum proofs}

An area of theoretical computer science that is very important and
related to complexity is the field of {\em proofs}. The concept of
short classical proofs for a statement is captured by the complexity
class~$\NP$ and is the most natural. We know that many difficult
problems actually have short witnesses or proofs. Can we generalize
this concept in a useful and meaningful way to the quantum world?
What would be a quantum proof? Would it be useful?

In a seminal paper by John Watrous \cite{Watrous00}, a specific
problem, group non-membership, was shown to have short quantum
proofs. It is not known (and believed to be impossible) to come up
in general with a short classical proof that an element is not part
of a group when the description of the group is given as a list of
generators. What is amazing is that there exist quantum states that
can be associated to such a problem that are short quantum proofs.
More specifically, if the verifier has a quantum computer, there is
a quantum algorithm that will efficiently verify the witness: if the
element is in the group, no quantum state will make the verifier's
algorithm accept with non-negligible probability, whereas if the
element is not in the group, there is a quantum state that will make
the algorithm accept with probability~1.

\subsection*{Classical simulation of entanglement}

In previous sections, we presented several examples where
entanglement can be used to solve distributed computing problems
more efficiently. In physics and computer science, an active area of
research is dealing with the opposite problem, the simulation of
entanglement using classical communication. The objective is to
exactly  reproduce the distribution of measurement outcomes, as if
they were performed on entangled qubits. 
We assume in what follows that the distant parties share an infinite amount of random variables, because otherwise it is impossible to simulate quantum correlations with worst case communication~\cite{MBCC01}.
 The first protocol to simulate a maximally
entangled pair of qubits using classical communication was presented
in \cite{Maudlin}. The protocol uses an expected 1.74 bits of
communication but to be able to simulate a maximally entangled pair
of qubits perfectly, the amount of communication is not bounded. In
\cite{BCT99}, a simulation was presented using exactly 8 bits of
communication and this was later improved to 1
bit~\cite{TonerBacon}.

In general, looking at the classical communication complexity (with
shared randomness) for pseudo-telepathy games tells us  how
difficult it is to simulate entanglement. Using this idea, it is
proved in~\cite{BCT99} that $n$~maximally-entangled qubits require
an exponential amount of communication for  perfect simulation.
Some protocols actually exist that accomplish this almost tightly
with an expected amount of communication  for \emph{general} measurements~\cite{MBCC01}. \looseness-1

\subsection*{Leader election}
An important problem in distributed computation is \emph{leader election}. The task is the following: within a group of distributed parties that communicate according to an underlying graph structure, create the situation where a single participant is singled out as the \emph{leader}. In~\cite{TKM05},  Seiichiro Tani, Hirotada Kobayashi, and Keiji Matsumoto give an error-free protocol for the leader
election problem in the anonymous networks setting.
This contrasts with the fact that any classical bounded-time protocol must have a non-zero error probability.
See also~\cite{DP06}.

\subsection*{Byzantine agreement}

In \emph{Byzantine agreement}~\cite{LSP82} (also called \emph{consensus}), a group of participants must decide on a common output bit value, despite the intervention of an adversary. The participants each have an input bit. The protocol must be such that all honest participants agree on the same output bit, and if the inputs were all~$c \in \{0,1\}$, the common output bit is~$c$. The complexity of the protocol is measured by the maximal expected number of rounds it takes for the protocol to terminate. In~\cite{BH05}, Michael Ben-Or and Avinatan Hassidim present a fast quantum Byzantine agreement protocol that can reach agreement in $O(1)$ expected communication rounds. This beats the  known classical lower bound of $O(\sqrt{n/\log n})$ and the~$O(n)$ complexity of best known
classical solutions.

\subsection*{Protocols for quantum information}

If we choose to deal with tasks involving quantum information
instead of classical information, there are a lot of results and
possibilities. Quantum teleportation is the most
famous~\cite{teleport}, but all sorts of channels have been studied
for quantum communication. On the cryptography side, we know
protocols to encrypt~\cite{PQC} and authenticate quantum
messages~\cite{BCGT02}. It is possible to perform multi-party
computation with quantum inputs and outputs in a secure
way~\cite{BCGHS06}. It is also possible to anonymously transmit
quantum messages~\cite{BBFGT07}.

\section*{Conclusion}

We have given the reader a glimpse of distributed computing in the
quantum world. Following the main lines of our survey, we now
present a partial list of open questions.

\emph{Characterization of games that exhibit pseudo-telepathy.} One
way to recognize a pseudo-telepathy game is to find a perfect
quantum strategy and then show that there is no such classical
strategy. We would like a more natural way to recognize such a game,
relying more on its underlying structure.

\emph{Quantum parallel repetition.} What is the best probability of
success for Alice and Bob who are involved in many \emph{parallel}
instances of the same game, using entanglement? For purely classical
games, the probability of success decreases at an exponential
rate~\cite{Raz98} (as surprising as it sounds, the probability does
not decrease at the same rate as one might expect and this result is far from being
trivial).  This question asks whether or not there is a similar
theorem for the case that the players use shared entanglement. A
special case was answered in the affirmative by~\cite{CSUU07}.

\emph{Quantum communication complexity: qubits versus entanglement.}
As mentioned, we know that teleportation can be used to transform
any two-player protocol using quantum communication into a protocol
using entanglement, at a cost of only two classical bits per qubit
in the original protocol. This question asks whether or not we can
do the same thing, up to a constant factor, in the \emph{other}
direction. Related work in this direction includes~\cite{LTW08},
where it is shown that in a slightly different scenario, there exist
tasks for which no finite amount of entanglement yields an optimal
strategy.

\emph{Simulation of multi-party entanglement.} In contrast to the
two-party case, very little is known about the simulation of
multi-party entangled states. In particular, it is not even known if
this general task is possible with bounded communication.

\section*{Acknowledgements}

We thank John Watrous for  related  insightful discussions and Michael Ben-Or for pointing out references on quantum leader election and Byzantine agreement.

\bibliographystyle{plain}
\bibliography{references}

\end{document}